# Baryons in the Constituent-Quark Model

David Akers*
*Lockheed Martin Corporation, Dept. 6F2P, Bldg. 660, Mail Zone 6620,*
*1011 Lockheed Way, Palmdale, CA 93599*
*Email address: David.Akers@lmco.com

**Abstract**

An elementary constituent-quark (CQ) model of mesons was previously presented. In this paper, we continue research into a study of the baryons in the constituent-quark model. Mac Gregor proposed a comprehensive model of elementary particles for which both mesons and baryons shared common mass-band structure in quantized units of m = 70 MeV, B= 140 MeV and X = 420 MeV. A review of the baryon data is under taken for comparison with the CQ model. It is shown in this paper that baryons possess an isospin *I* related to the mass quantum m = 70 MeV and to the B = 140 MeV quantum (or the mass of the pion). In order to establish a consistency with the quark model of Gell-Mann, we identify the SU(3) baryon decuplet as a standard feature to be maintained with only slight changes to the constituent-quark masses. By insisting on the J = 3/2, P-states of the SU(3) baryon decuplet to be in the *same* CQ excitation states, we are lead to establish baryon cores in the P-states with J = 1/2. Core corrections to Mac Gregor's CQ model of baryons are presented. Exact shell structure is found among all the baryons regardless of isospin as evidenced in the data from the Particle Data Group listing. New baryons are predicted to exist. The possible existence of magnetic charge in hadronic structure is suggested.

## 1 Introduction

Although quantum chromodynamics (QCD) has shown tremendous success in describing particles physics over the last few decades [1-3], there are a few fundamental problems which still remain: namely, what is the origin for the masses of elementary particles? What is the source of the spin of the nucleon? Where is the predicted magnetic monopole of Dirac? In a parallel development to QCD, there have been proposals by Schwinger and others that quarks may consist of both electric and magnetic charges [4-11]. The present paper on the baryon spectrum is a sequel to an earlier paper on the meson spectrum [12], where it was suggested that magnetic charge of spin J = 0 may exist internal to the quarks and generate magnetic fields among the meson states. The existence of Zeeman splitting among the meson states was presented as possible evidence. Moreover, Sawada has suggested evidence for magnetic charge in scattering experiments to account for residual strong interaction effects [10]. In Ref. [12], the existence of a 70 MeV boson was derived from the mass of the classical Dirac magnetic monopole. Evidence for the existence of a 70 MeV boson has been known for some time [13-14]. We present additional evidence for the energy scale of 70 MeV in the present work.

In a study of the mesons and baryons, Mac Gregor [15] developed a comprehensive constituent-quark (CQ) model of elementary particles. It was shown that the CQ masses are directly related to the masses of the electron, muon, and pion. A connection was later discovered between the CQ masses of Mac Gregor's model and Nambu's empirical mass formula $m_n = (n/2)137m_e$, $n$ a positive integer and $m_e$ the mass of the electron [11]. Mac Gregor's esoteric notation included a 70 MeV quantum, a boson excitation B with the mass of the pion at 140 MeV, a fermion excitation F with a mass of 210 MeV or twice the muon mass, and a 420 MeV excitation quantum X. The 70 MeV quantum and the 420 MeV quantum X do not correspond to any observed particles but serve as the building blocks of mesons and baryons in the CQ model. The mechanism for generating the CQ masses is discussed in Refs. [12, 15] and will be briefly discussed in Section 4. It is sufficient to say that we shall accept the evidence of Mac Gregor's CQ model from its previous agreement with earlier experimental data and that we shall present further evidence for the 70 MeV excitation quantum from the Particle Data Group listing [16]. The experimental evidence for meson spectra in the CQ model was previously presented [12]. For the purpose of our study, we review the common mass-band structures of the baryons. This structure was tabulated in Mac Gregor's work as Table XIX, which can be described as the periodic table of the baryons [15]. It is this table which we update in the present work.

In the CQ model, the mass of a resonance is determined mainly from the masses of the constituent quarks. Spin and orbital excitations can also contribute to the mass of the resonance with higher mass-states appearing at higher total angular momentum values. However, both baryons and mesons with different J-values appear in the same mass-band structure in Fig. 1 of Mac Gregor's work [15]. These meson resonances appear with accurate $J \sim M^2$ Regge trajectories for $J^P = 1^-$, $2^+$, and $3^-$. Likewise, the meson spectrum also exhibits non-Regge spacing with $J \sim M$ for the $2^+$, $3^-$, $4^+$, $5^-$ and $1^+$, $2^-$, $3^+$ yrast levels in Fig. 2 of Mac Gregor's paper [15]. The spacing of the Regge-like and non-Regge levels is in accurate 420 MeV intervals, for which Mac Gregor assigned an excitation quantum $X^1$ = 420 MeV. The quantum X can appear with zero or non-zero units of angular momentum.

Additional sets of baryon resonances are found to depend upon their quark content. For the CQ model, the constituent-quark basis states are calculated to be u(315), d(320), s(525), c(1575), and b(4725). To maintain correct mass values for the baryon SU(3) decuplet, we utilize instead u'(385), d'(395), and s'(595). The CQ model was postulated long before the advent of the top quark and has not been extended to the energy range for resonances involving the top quark. For purposes of our study, we limit our discussion to baryons below 2700 MeV. As shown in Fig. 4 of Mac Gregor's paper [15], there are baryon resonances equally spaced by the excitation quantum B = 140 MeV. Mac Gregor concluded that the excitation quantum B served as a fundamental mass unit.

## 2 Baryon Masses

In this paper, it is shown that there is exists a mass-band structure in units of m = 70 MeV, separating baryon states at high angular momentum J, and that there exists a common shell structure for baryons of different isospins I. This evidence is shown in Table 1, which has the usual CQ model notation. In Table 1, we list all the well-established baryons from the Particle Data Group [16]. The not-so-well-established baryon resonances are also included in the table. The vertical columns in Table 1 represent increasing energy in quantized units of mass m = 70 MeV for baryons in radial-, orbital- and spin-space. The horizontal rows represent increasing energy in quantized units of m = 70 MeV for baryons in isospin-space. As shown in Table 1, shell numbers are shown to exist in isospin space as well and are in units of m = 70 MeV. We have introduced baryon core corrections in Table 1 to be explained. Evidence for the existence of a baryon core is well known [17-19].

For the Δ baryon resonances, Mac Gregor utilized the nucleon N(939) as the ground state. By the selection of the N(939) core for the Δ baryons, there is an inconsistency with the Gell-Mann quark model in which the SU(3) decuplet for J = 3/2 has equally spaced mass separations of approximately 140 MeV. Therefore, the core selection was wrong for the Δ baryons in Ref. [15]. Likewise, we note that the core selection by Mac Gregor for the Ω baryons was wrong for the same reasons as stated in regards to satisfactorily explaining the masses of the SU(3) decuplet. Mac Gregor chose to list the Δ baryons after the (Δ - N) core correction and chose to place the Ω(1672) in the ground state with J = ½. However, Ω(1672) has a J = 3/2 and should be located in the *same* CQ excitation band as the other members of the SU(3) decuplet. The SU(3) decuplet is noted as the underlined baryons in row F(210) of Table 1. Thus, there must exist a Ω(1499) core with spin J = ½ which is not previously known.

The Δ(1232)P is now located in row F(210) of Table 1 for the reasons previously explained. The baryons in row F(210) all have spin J = 3/2 in the P-state. This establishes a consistency with the Gell-Mann quark model. However, there remains what to select for the Δ baryon core. It is obvious from the ground states across all isospins that the Δ core must have J = ½ in the P-state and that it must have a mass of about 1079 MeV. With the Δ(1079)P core correction, we introduce a new column of Δ baryons in Table 1, where all baryons have the *same* J-value in each row. Moreover, we have a new column of Ω baryons in Table 1 as well. In Table 1, the baryons in boldface were predicted in Ref. [13], those baryons in blue color indicate rotational states by Mac Gregor, and the author's predictions of baryon spins are indicated in red color.

## 3 Baryon Excitations and Meson States

In order to understand the baryon excitations of Table 1, we first must identify the energy scale, so that we can make comparisons between the baryons and the mesons. We utilize a scale of particle masses based upon the CQ model as found in Fig. 3 of the paper by Mac Gregor [15]. In fact, there are two distinct scales in the figure; one scale starts with the pion mass at 140 MeV and has steps of X = 420 MeV, and the other scale starts at zero and has steps of q = 315 MeV. The X = 420 MeV scale has particle masses at $\pi(140)$, $\eta(547)$, $\eta'(958)$, $\eta(1440)$, $\eta(1760)$, and $\eta(2225)$. The q = 315 MeV scale has particle masses at $\eta(1295)$, $\eta(1580)$, and D(1864); these particles can be identified at the mass levels 4q = 1260 MeV, 5q = 1575 MeV, and 6q = 1890 MeV, respectively. These levels can be rearranged as follows: 4q = 3(420) = 630 + 630, a triple and transform reaction as noted in Ref. [15], so that the mass scales are interchangeable between mesons and baryons under the appropriate rules. 5q is more problematic, because this would seem to suggest that the meson $\eta(1580)$, a boson, is in fact a fermion from the quark content. However, with the appropriate binding energy rules in Ref. [14], the predicted meson $\eta(1580)$ seems to fit better into the 1540 or 1610 MeV levels of Table 2 in Ref. [12]. Binding energies of the composite mesons are also discussed in Mac Gregor's work [15]. Finally, the D(1864) is easily identified by the rearrangement of the quark content as follows: c = 5q = 1575 and u = q = 315; thus, D(1864) = c q, where c is the charm quark and q is the u or d antiquark.

The set of mesons corresponding to the 420 MeV scale is shown in Fig. 1. In Fig. 1, the experimental meson masses are indicated by solid lines and are taken from the Particle Data Group [16]. The vertical arrows represent energy separation of about 420 MeV between states. The X = 420 MeV is the spin-orbit energy separation between the singlets in the S-states and the P-states. We note a consistent pattern of energy separation between the spin-singlet and -triplet states. The lowest lying charmonium states are also shown for comparison, and the X = 420 MeV energy of separation is indicated by the arrows. There are distinct groupings as indicated by the arrows. The $\eta(2980)$ is associated with the $\chi_{c0}(3415)$ state from extensive study of the charmonium spectrum. Taking this pattern of energy separation to the lower meson states, we note that there are associated groupings or meson partners. $\eta(1295)$ is associated with $f_0(1710)$, $\eta'(958)$ with $f_0(1370)$, and $\eta(547)$ with $f_0(980)$. For the lowest lying state, $\pi(140)$ has a missing associated partner. A missing $f_0$ meson is shown at 560 MeV and is predicted to exist.

The $f_0(560)$ is a missing partner of the pion in the CQ model as inferred from Fig. 1. In the study of mesons in this mass range, there has been extensive debate in regards to the existence of the $\sigma(400\text{-}1200)$ meson. Numerous models have predicted the existence of the $\sigma$ meson, and in fact this meson is now identified as the $f_0(400\text{-}1200)$ scalar [20] and listed as the $f_0(600)$ by the Particle Data Group [16]. Van Beveren *et al* identified this scalar meson as the dynamically generated chiral partner of the pion [20]. It is interesting that the $f_0(560)$ is easily identified as a missing partner of the pion in Fig. 1.

The baryon spectrum, *without* core corrections, can be plotted as shown in Fig. 2. In Fig. 2, there is no obvious pattern of energy separation between states. The meson scale is shown in the bin with spin J = 0 for mass comparisons only. The baryon spectrum of Table 1, *with* core corrections, can now be combined with a few of the low-mass mesons of Fig. 1 and shown for mass comparisons only. For each isospin, core masses (with J = ½) are subtracted from each baryon resonance in the vertical columns of Table 1. The differences are then plotted in Fig. 3. We must emphasize that the mesons and baryons do *not* have the same J-values. The results of these mass comparisons are shown in Fig. 3. The J = 1 mesons are overlapped with the J = ½ baryons for mass-scale comparisons only. In Fig. 3, the mesons are indicated by black lines and the baryons by color: N(green), Δ(blue), Λ(pink), Σ(red), Ξ(purple), and Ω(yellow). The black-dashed lines represent baryons, which are predicted to exist. In Fig. 3, there is noted some evidence of precise mass quantization (m = 70 MeV) in the high angular momentum *L*- values (total J = *L* + *S*). We emphasize here that the mass differences in Fig. 3 are derived from the experimental data [16] and that the m = 70 MeV spacings are real. These spacings can be compared to the same accurate spacings found in Fig. 3 of Mac Gregor's work in Ref. [13].

In the high angular momentum states, we see clear evidence of mass quantization in units of m = 70 MeV. Three levels (or F = 210 MeV) are indicated by the vertical arrow at the right in Fig. 3. At the high angular momentum states near spin J = 9/2, there appears to a pattern of shell closure. Several authors have suggested the idea of particle shells [21-23]. In the region for relatively high L-values (or J-values), the baryons appear to be linearly separated in vibration or excitation states. At the low angular momentum region, there appears to be band structure as shown Figs. 3 and 5. However, the experimental uncertainties in mass must be reduced to clarify the situation for masses near 700 to 840 MeV and for spin J = 5/2. The Ω(1672) baryon, after core subtraction, appears to be in one of the four equally separated states at the 140 MeV level. We indicate the π(140), K(494), N(939) and Ω(1672) masses on the same energy scale in Fig. 3. These particles have lifetime stability peaks [21], and they are found with lifetimes in low quantized powers (n = - 4, 1, 1, 2, respectively) of the fine-structure constant α = 1/137 [24].

## 4 Physics Beyond the Standard Model

We are now in a position to ask what is the physics involved for the baryons of Fig. 3. Although the Standard Model of elementary particles does not have the features of the particle spectrum indicated in Fig. 3, we can incorporate some of the known laws of physics to possibly explain the baryon and meson spectra. How do the meson states, with high angular momentum, compare to the baryon spectrum of Fig. 3? A set of mesons corresponding to the 420 MeV scale is shown in Fig. 4. In Fig. 4, there is a consistent pattern of spin-spin and spin-orbit energy separation between the states in comparison to the charmonium states [12]. The spin-orbit interactions are about 420 MeV and the spin-spin interactions are on the order of 35 to 100 MeV. The set of particles η'(958), ϕ(1020), $f_0$(1370), $f_1$(1465), and $f_2$(1525) parallel the lowest lying charmonium states [12]. In Fig. 4, the solid lines represent the well-established particles taken from the Particle Data Group [16]. The dot-dashed lines represent mesons, which are predicted to exist from the symmetry of the pattern. For all figures in this paper, the solid lines represent experimental meson masses, and the dashed or dashed-dot lines represent unobserved particles, which are predicted to exist.

We can now overlap the meson states of Fig. 4 with the baryon spectrum of Fig. 3. The spectra are combined for purposes of mass comparisons only. This result is shown in Fig. 5. The color code of the particles is the same as in Fig. 3. The overlap of meson and baryon masses is indicated in the high angular momentum states. Again, we must emphasize that mesons and baryon do *not* have the same J-values. We note that the precise energy separations in the baryon states exceed those found in the meson states. This may be due to experimental uncertainties in the meson masses and to the fact that large baryon cores are subtracted.

In Ref. [12], we postulated the possible existence of magnetic charge in hadronic structure, which followed from Schwinger's idea of dyonic quarks [4-5] and from Chang's suggestion [6] that quarks may consist of both electric and magnetic charges in describing baryons. Dyons conserved angular momentum [5]. Thus, we can also list some possibilities from the laws of physics for the spectrum in Fig. 5; namely, we could have the following: 1) Coulomb interactions and Exchange Coulomb interactions, 2) weak Zeeman Effects or strong Paschen-Back Effects, 3) Stark Effects, and 4) rotational-vibrational spectra. The idea of rotational spectra in particle physics was suggested by Mac Gregor [13]. For the first, Coulomb interactions would result in non-linear energy separation in the spectra. The second of these interactions, namely the strong Paschen-Back Effect has been suggested for the meson spectra [12]. It was noted in [12] that there is possible evidence for Russell-Saunders or *LS* coupling in the meson states. These meson spectra can be shown to satisfy the Lande interval rule, which is widely used in atomic, molecular and nuclear physics. A more extensive study of the Lande ratio for the P-states would involve a complete derivation of the Lande factor g in the Zeeman or Paschen-Back energy splitting of these meson states:

$$\Delta E = - \mu_B \, g \, M_J \, B. \qquad (1)$$

Zeeman or Paschen-Back splitting would allow linear energy separation between the particle states [25, 26]:

$$\Delta E = 2K\,(j + 1). \qquad (2)$$

Eq. (2) is the separation in the energy of adjacent levels of a multiplet and is proportional to the total angular momentum quantum number of the level of higher energy [26]. Eq. (2) is called the Lande interval rule. In Eq. (1), the magnetic field could be generated from the presence of magnetic charge internal to the quarks. The individual quark's magnetic moment would then interact with the field B. Typical magnetic interactions are the order of 50 MeV, depending upon the rms radius of the charge distribution for the mesons ($r \sim 0.6$ fm) compared to that for the baryons ($r \sim 0.8$ fm) [27].

If we consider the third possibility of the Stark Effect, there could be energy shifts, which are linear, in a uniform electric field. It is possible that the electric dipole moment of baryon core interacts with the electric field of an orbital quark. However, the field of a point charge is radial and non-uniform. Therefore, the energy shifts would not be expected to be linear in such a situation. However, if the electric and magnetic field lines are confined in a flux tube, then there may be possible QCD effects.

Finally, we come to the fourth idea of vibration-rotational spectra [13]. If we consider the mesons and baryons as simple systems of classical vibration-rotation, then there is linear energy separation between adjacent particle states. The energy of a vibrational motion can be quantized:

$$E_n = (n + \tfrac{1}{2})\, h\nu. \qquad (3)$$

In the J = 3/2 bin, the energy between adjacent nucleons (green) in Fig. 3 is about F = 210 MeV. This is about 22% the mass of the N(939). In the J = ½ bin, the energy separation is about m = 70 MeV or 7% the mass of the N(939). However, we must remember that baryon cores are subtracted in Figs. 3 and 5 in order to calculate any physics effects.

On the other hand, for rotational energy with a classical center-of-mass, the energy can be quantized:

$$E_n = (1/2I)\,(h/2\pi)^2\,[n(n + 1)], \qquad (4)$$

where I is the moment of inertia, h is Planck's constant, and n = 0, 1, … The two-body (meson) or three-body (baryon) could be treated as simple molecules with rotational energy. In fact, we note several bands of closely spaced baryons in Fig. 5. For the J = 3/2 bin, we have the following sets of band-mass separations:

a) $[\Xi(1531.8) - \Xi_{core}(1314.83)] - [\Sigma(1382.8) - \Sigma_{core}(1189.37)] = 23.54$ MeV;

b) $[\Sigma(1382.8) - \Sigma_{core}(1189.37)] - [\Omega(1672.45) - \Omega_{core}(1499)] = 20$ MeV;

c) $[\Omega(1672.45) - \Omega_{core}(1499)] - [\Delta(1232) - \Delta_{core}(1079)] = 20.45$ MeV.

These sets of band-mass separations are suggestive of a 20 MeV quantum, and they may represent evidence of possible rotational energy or even internal structure effects [28]. This energy separation is smaller than the quantum m = 70 MeV as noted in Figs. 3 and 5. For the missing baryons indicated by the dashed lines in Fig. 3, we calculate the possible masses and spins of these baryons in Table 2. Many of these predicted baryons may not exist for dynamical reasons; however, the symmetry of the pattern in Fig. 3 suggests that a few of the baryons may exist to fill in the missing states.

## 5 Conclusion

In this paper, an elementary constituent-quark (CQ) model of baryons was presented. Mac Gregor proposed a comprehensive model of elementary particles for which both mesons and baryons shared common mass-band structure. The existence of a 70-MeV quantum was postulated by Mac Gregor and was later shown to fit the Nambu empirical mass formula $m_n = (n/2)137m_e$, $n$ a positive integer. A review of the baryon data was under taken in this paper for comparison with the CQ model. It was shown that baryons possess an isospin *I* related to the mass quantum m = 70 MeV and to the B = 140 MeV quantum (or the mass of the pion). By insisting on the J = 3/2, P-states of the SU(3) baryon decuplet to be in the *same* CQ excitation states, we were led to establish baryon cores in the P-states with J = 1/2. Core corrections to Mac Gregor's CQ model of baryons were presented. Exact shell structure was found among all the baryons regardless of isospin. The existence of new baryons was predicted from the symmetry of the patterns as shown in Figs. 3 and 5.

The study of baryon and meson decays will lead the way to resolve which of the possible theories presented here is correct, and further experimental efforts will identify the existence or non-existence of the predicted mesons and baryons.

***Note Added in Proof.***

Evidence for the presence of magnetic charge or monopoles is being currently studied by experimentalists [29]. A current bibliography on magnetic monoples may be found in [30]. Since the writing of this paper, there have been a number of discoveries by various groups. First, there is reported evidence for low-mass ( < 1460 MeV) baryons [31]. This is consistent with the prediction of new baryons in Tables 1 and 2. However, experimental confirmation remains to be seen. Second, the LEPS Collaboration [32] has discovered a S = +1 baryon resonance at 1540 MeV, which has been confirmed by the CLAS Collaboration [33]. In Section 3, we noted that the 5q = 1575 MeV state would be problematic for the η(1580) meson and that this meson would seem to fit better into the 1540 or 1610 MeV levels of Table 2 in Ref. [12]. In the CQ model, the nucleon quark mass is q = 315 MeV. With consideration of 2-3% binding energies, the pentaquark state would be a little less than 5q = 1575 MeV and would be consistent with the discovery of the S = +1 baryon resonance at 1540 MeV. Mac Gregor noted long ago that both mesons and baryons share the same constituent quark mass bands. Finally, the author discovered in a literature search a quantum-mechanical derivation of the Zeeman effect for an electric charge – magnetic monopole system by Barker and Granziani [34]. This derivation is shown in their Eqs. (35S) and (36M) in Ref. [34]. Thus, the idea of Zeeman splitting is possible in hadron spectroscopy in the presence of magnetic charge.

## Acknowledgement

The author wishes to thank Dr. Malcolm Mac Gregor of the University of California's Lawrence Livermore National Laboratory for his encouragement to pursue the CQ Model, and he wishes to thank Dr. Paolo Palazzi of CERN for his interest in the work and for e-mail correspondence.

Table 1. A constituent-quark (CQ) mapping of all the baryon resonances is listed for < 2700 MeV from the Review of Particle Properties [16]. Core corrections of the J = ½ ground states are made for the Δ and for the Ω baryons, which were not done by Mac Gregor in 1990 [15]. Underlined baryons indicate SU(3) decuplet for J = 3/2. Baryons in boldface were predicted in Ref. [13]. Baryons in blue indicate rotational states. The author's predictions of baryon spins are indicated in red.

| | | | | | | |
|---|---|---|---|---|---|---|
| Isospin | ½ | 3/2 | 0 | 1 | ½ | 0 |
| Shell numbers (on 70 MeV scale) | N | N + 2m | N + 3m | N + 4m | N + 6m | N + 8m |
| Ground state | N | NB | Λ = NF | Σ = NBB | Ξ = NBBB | Ω = NBBBB |
| (in MeV) | (939) | Δ(1079)P | Λ(1116)P | Σ(1192)P | Ξ(1321)P | Ω(1499)P |
| CQ excitation (in MeV) | | | | | | |
| m(70) | | **Δ(1149)S** | | | | |
| B(140) | | | | | | |
| F(210) | | Δ(1232)P | **Λ(1326)** | Σ(1385)P | Ξ(1530)P | Ω(1672)P |
| BB(280) | **N(1219)** | | Λ(1405)S | Σ(1480)$S_{1/2}$ | | |
| FB(350) | | | | Σ(1560)$P_{1/2}$ | Ξ(1620)$P_{1/2}$ | |
| X(420) | **N(1359)** | | Λ(1520)D | Σ(1580)D | Ξ(1690)$D_{3/2}$ | |
| FBB(490) | **N(1440)P** | | Λ(1600)P | Σ(1620)S | | |
| | | | Λ(1670)S | Σ(1660)P | | |
| | | | | Σ(1670)D | Ξ(1820)D | |
| BX(560) | | Δ(1600)P | Λ(1690)D | Σ(1690)$P_{3/2}$ or $D_{3/2}$ | | |
| | | Δ(1620)S | | Σ(1750)S | | |
| | | | | Σ(1770)P | | |
| | | | | Σ(1775)D | | |
| | | | | | Ξ(1950)? | |
| FX(630) | **N(1520)D** | Δ(1700)D | | Σ(1840)P | | |
| | N(1535)S | Δ(1750)P | Λ(1800)S | Σ(1880)P | | |
| BXB(700) | N(1650)S | | Λ(1810)P | | | |
| | N(1675)D | | Λ(1820)F | Σ(1915)F | Ξ(2030)$D_{5/2}$ or $F_{5/2}$ | |
| | **N(1680)F** | | Λ(1830)D | | | |
| | N(1700)D | | | Σ(1940)D | Ξ(2120)$D_{3/2}$ | Ω(2250) $D_{3/2}$ |
| FBX(770) | N(1710)P | | | Σ(2000)S | | |
| | N(1720)P | | Λ(1890)P | | | |
| XX(840) | | Δ(1900)S | | | | |
| | | | | Σ(2030)F | | |
| | | Δ(1905)F | | Σ(2070)F | | |
| | | Δ(1910)P | | | | |
| | | Δ(1920)P | | Σ(2080)P | | |
| | | Δ(1930)D | | | | |
| | | Δ(1940)D | | | | |
| | | Δ(1950)F | | | | |
| mXX(910) | | | Λ(2000)$G_{7/2}$ | Σ(2100)G | Ξ(2250)$G_{7/2}$ | Ω(2384)$G_{7/2}$ |
| | | Δ(2000)F | | | | |
| | | | Λ(2020)F | | | |
| BXX(980) | N(1900)P | | | | | Ω(2470)$P_{3/2}$ |
| | | | Λ(2100)G | | | |
| | | | Λ(2110)F | | Ξ(2370)$F_{5/2}$ | |
| FXX(1050) | N(1990)F | | | Σ(2250)? | | |
| | N(2000)F | | | | | |
| | | Δ(2150)S | | | | |
| $B_2X_2$(1120) | N(2080)D | | | | Ξ(2500)$D_{3/2}$ | |
| | N(2090)S | | | | | |
| | N(2100)P | | | | | |
| | | Δ(2200)G | | | | |
| $mB_2X_2$(1190) | | | Λ(2325)D | | | |
| XXX(1260) | N(2190)G | | | | | |
| | N(2200)D | | | | | |
| | **N(2220)H** | Δ(2300)H | | Σ(2450)$H_{9/2}$ | | |
| | | Δ(2350)D | | | | |
| | | | Λ(2350)H | | | |
| | | Δ(2390)F | | | | |
| $mX_3$(1330) | N(2250)G | Δ(2400)G | | | | |
| $BX_3$(1400) | | | | Σ(2620)? | | |
| $FX_3$(1470) | | Δ(2420)H | | | | |
| $X_4$(1680) | N(2600)I | | | | | |
| | N(2700)K | | | | | |

Table 2. Predicted masses and spins of new baryons for black-dash lines indicated in Fig. 3.

| Mass difference (in MeV) | Predicted Baryon | Mass (MeV) | Isospin I | Spin J |
|---|---|---|---|---|
| $N - N_{core} = 1071$ | N | 2010 | ½ | 3/2 |
| $\Delta - \Delta_{core} = 1071$ | Δ | 2150 | 3/2 | 3/2 |
| $\Lambda - \Lambda_{core} = 1071$ | Λ | 2187 | 0 | 3/2 |
| $\Sigma - \Sigma_{core} = 1071$ | Σ | 2263 | 1 | 3/2 |
| $\Xi - \Xi_{core} = 1071$ | Ξ | 2392 | ½ | 3/2 |
| $\Omega - \Omega_{core} = 1071$ | Ω | 2570 | 0 | 3/2 |
| $N - N_{core} = 1004$ | N | 1943 | ½ | 3/2 |
| $\Delta - \Delta_{core} = 1004$ | Δ | 2083 | 3/2 | 3/2 |
| $\Lambda - \Lambda_{core} = 1004$ | Λ | 2120 | 0 | 3/2 |
| $\Sigma - \Sigma_{core} = 1004$ | Σ | 2196 | 1 | 3/2 |
| $\Xi - \Xi_{core} = 1004$ | Ξ | 2325 | ½ | 3/2 |
| $\Omega - \Omega_{core} = 1004$ | Ω | 2503 | 0 | 3/2 |
| $N - N_{core} = 1205$ | N | 2144 | ½ | 5/2 |
| $\Delta - \Delta_{core} = 1205$ | Δ | 2284 | 3/2 | 5/2 |
| $\Lambda - \Lambda_{core} = 1205$ | Λ | 2321 | 0 | 5/2 |
| $\Sigma - \Sigma_{core} = 1205$ | Σ | 2397 | 1 | 5/2 |
| $\Xi - \Xi_{core} = 1205$ | Ξ | 2526 | ½ | 5/2 |
| $\Omega - \Omega_{core} = 1205$ | Ω | 2704 | 0 | 5/2 |
| $N - N_{core} = 1134$ | N | 2073 | ½ | 5/2 |
| $\Delta - \Delta_{core} = 1134$ | Δ | 2213 | 3/2 | 5/2 |
| $\Lambda - \Lambda_{core} = 1134$ | Λ | 2250 | 0 | 5/2 |
| $\Sigma - \Sigma_{core} = 1134$ | Σ | 2326 | 1 | 5/2 |
| $\Xi - \Xi_{core} = 1134$ | Ξ | 2455 | ½ | 5/2 |
| $\Omega - \Omega_{core} = 1134$ | Ω | 2633 | 0 | 5/2 |
| $N - N_{core} = 1191$ | N | 2130 | ½ | 7/2 |
| $\Delta - \Delta_{core} = 1191$ | Δ | 2270 | 3/2 | 7/2 |
| $\Lambda - \Lambda_{core} = 1191$ | Λ | 2307 | 0 | 7/2 |
| $\Sigma - \Sigma_{core} = 1191$ | Σ | 2383 | 1 | 7/2 |
| $\Xi - \Xi_{core} = 1191$ | Ξ | 2512 | ½ | 7/2 |
| $\Omega - \Omega_{core} = 1191$ | Ω | 2690 | 0 | 7/2 |

## Figure Captions

**Fig. 1.** Experimental meson masses, indicated as solid lines, are taken from the Particle Data Group [16]. The vertical arrows indicate the spin-orbit separation energy of about 420 MeV. The lowest lying charmonium states are shown for comparison. Note the consistent pattern of the separation energy between the spin-singlet and -triplet states. The $f_0(560)$ is predicted to exist.

**Fig. 2.** The total baryon spectrum, without core corrections, is combined with a few of the low-mass mesons of Fig. 1 for mass comparisons only (mesons and baryons do not have the same J-values). The mesons are indicated by black lines and the baryons by color: N (green), Δ(blue), Λ(pink), Σ(red), Ξ(purple), and Ω(yellow).

**Fig. 3.** The baryon spectrum of Table 1, with core corrections, is combined with a few of the low-mass mesons of Fig. 1 for mass comparisons only (mesons and baryons do not have the same J-values). The mesons are indicated by black lines and the baryons by color: N (green), Δ(blue), Λ(pink), Σ(red), Ξ(purple), and Ω(yellow). The black-dashed lines represent baryons, which are predicted to exist. There is noted evidence of precise mass quantization (m = 70 MeV) in the high angular momentum *L*- values (total J = *L* + *S*).

**Fig. 4.** The set of low meson masses associated with the η'(958) state. Experimental masses are indicated with solid lines and are taken from the Particle Data Group [16]. The black, dashed lines represent mesons, which are predicted to exist.

**Fig. 5.** The baryon spectrum of Fig. 3 and the mesons of Fig. 4 combined for mass comparisons only (mesons and baryons do not have the same J-values). The color code of the particles is the same as in Fig. 3. The overlap of meson and baryon masses is indicated in the J = 5/2 bin. The precise energy separations in the baryon states exceed those found in the meson states. This may be due to possible uncertainties in experimental measurements for the low-mass mesons and to the fact that large baryon cores are subtracted.

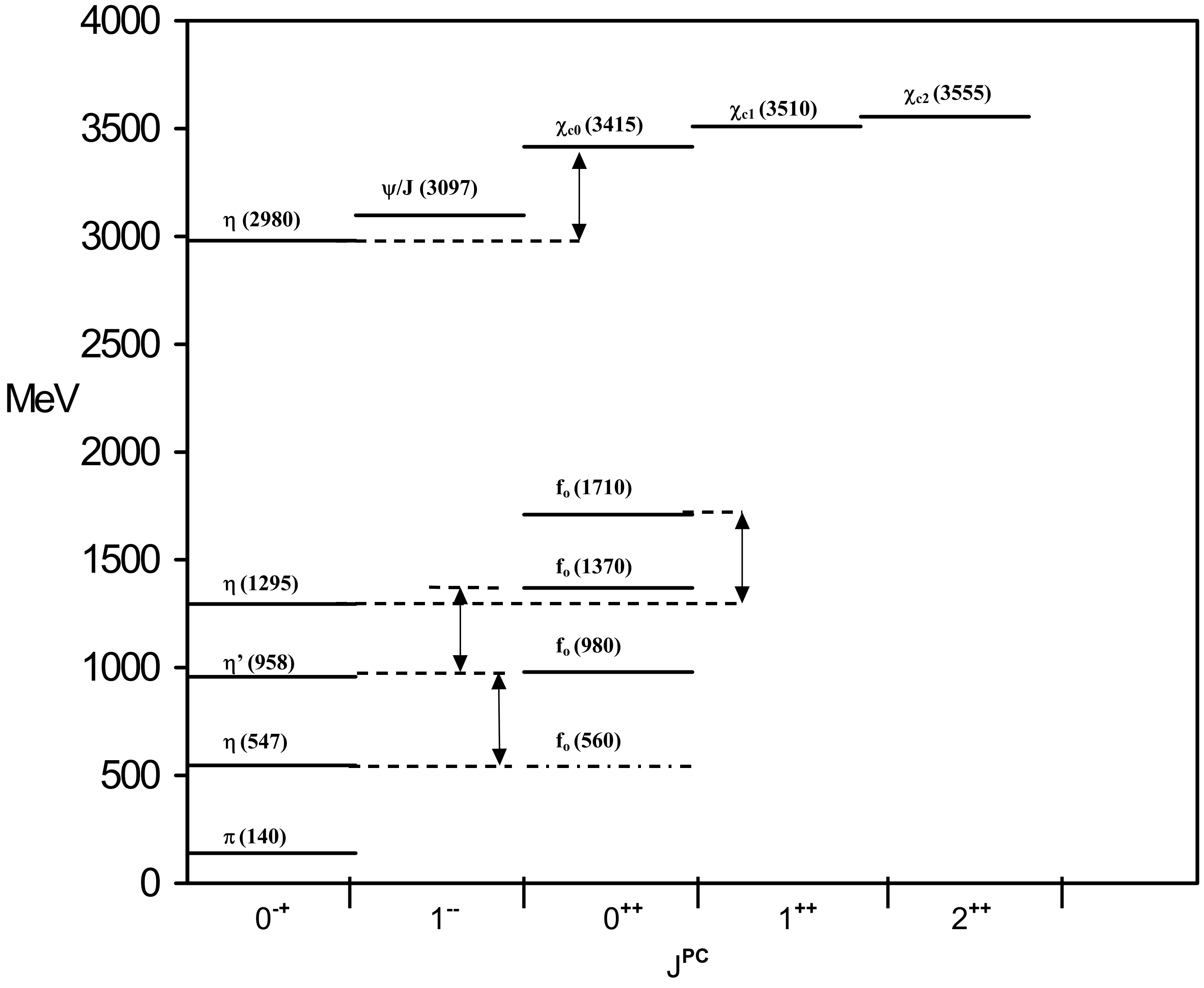

Fig. 1.

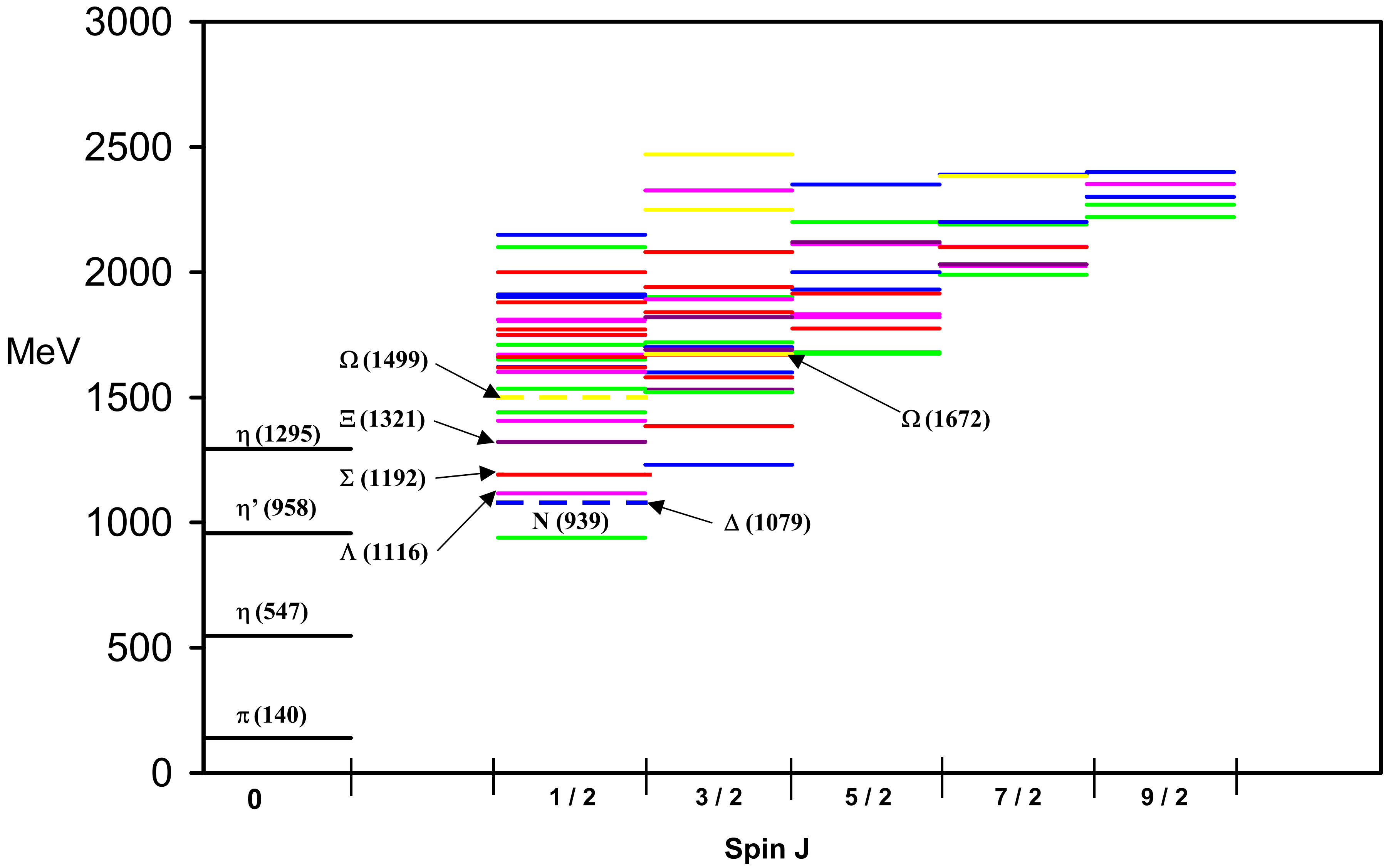

Fig. 2.

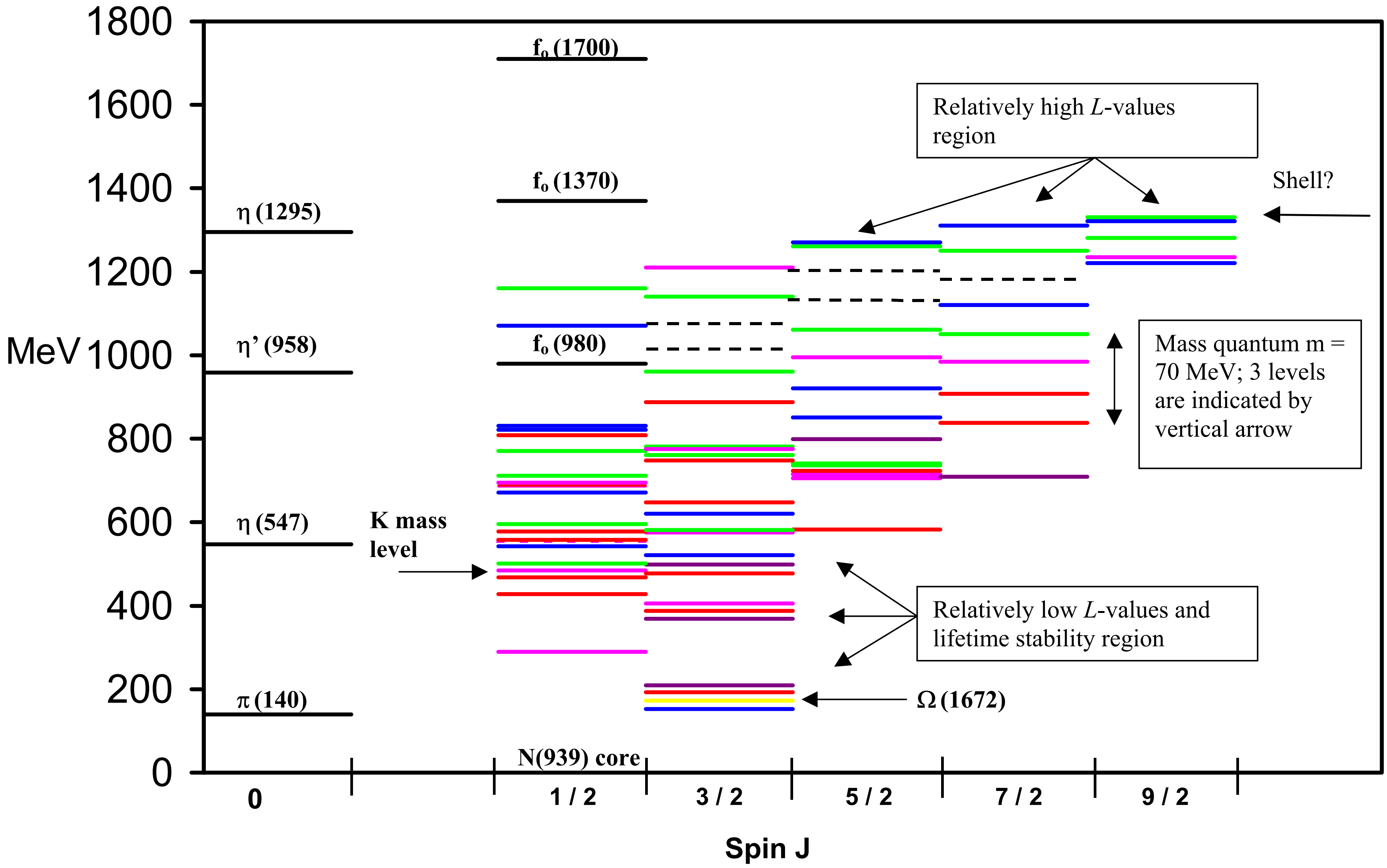

1800
1600
1400
1200
MeV 1000
800
600
400
200
0
f₀ (1700)
f₀ (1370)
f₀ (980)
η (1295)
η' (958)
η (547)
π (140)
K mass level
N(939) core
Relatively high L-values region
Shell?
Mass quantum m = 70 MeV; 3 levels are indicated by vertical arrow
Relatively low L-values and lifetime stability region
Ω (1672)
0
1 / 2
3 / 2
5 / 2
7 / 2
9 / 2
Spin J

**Fig. 3.**

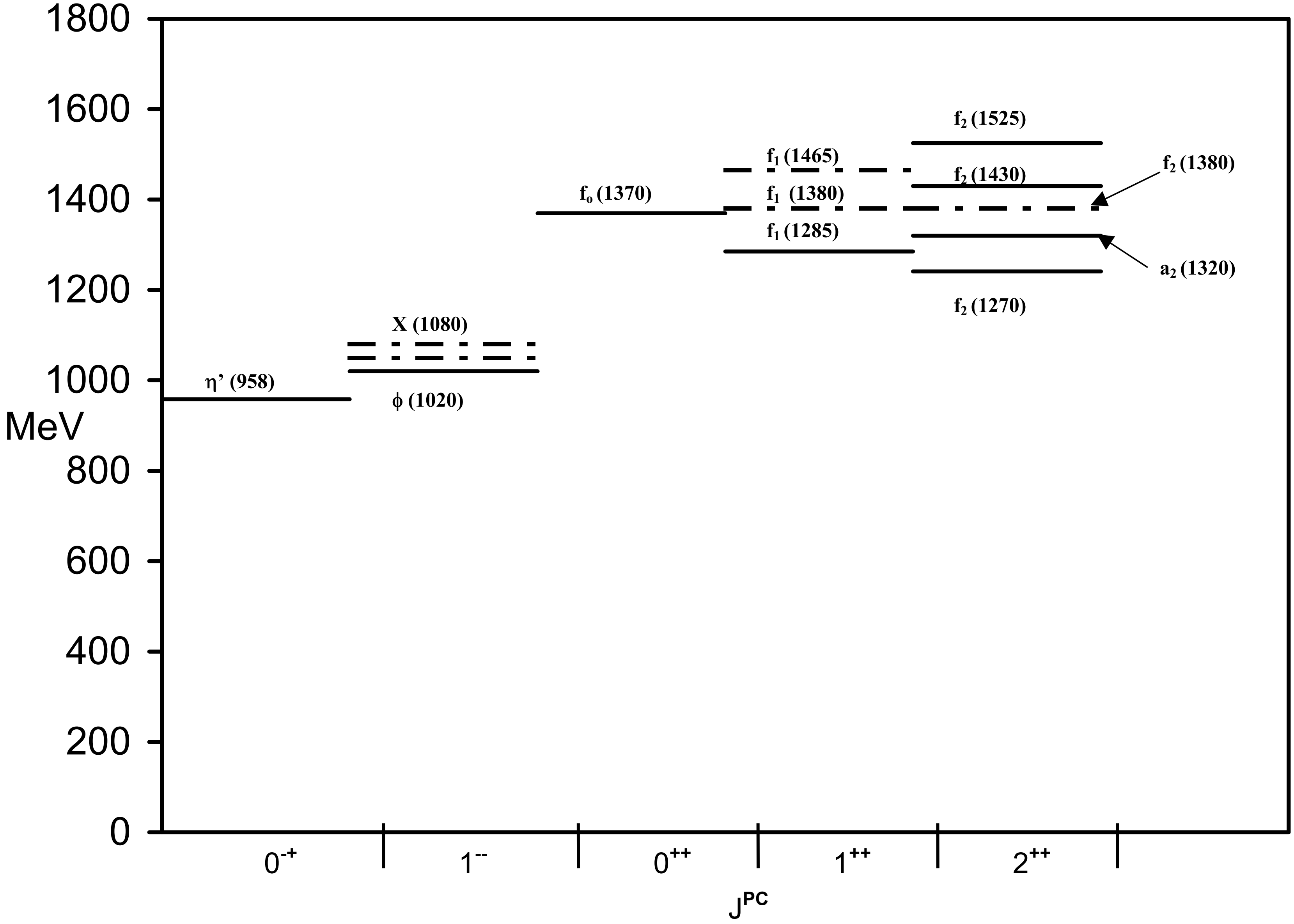

Fig. 4.

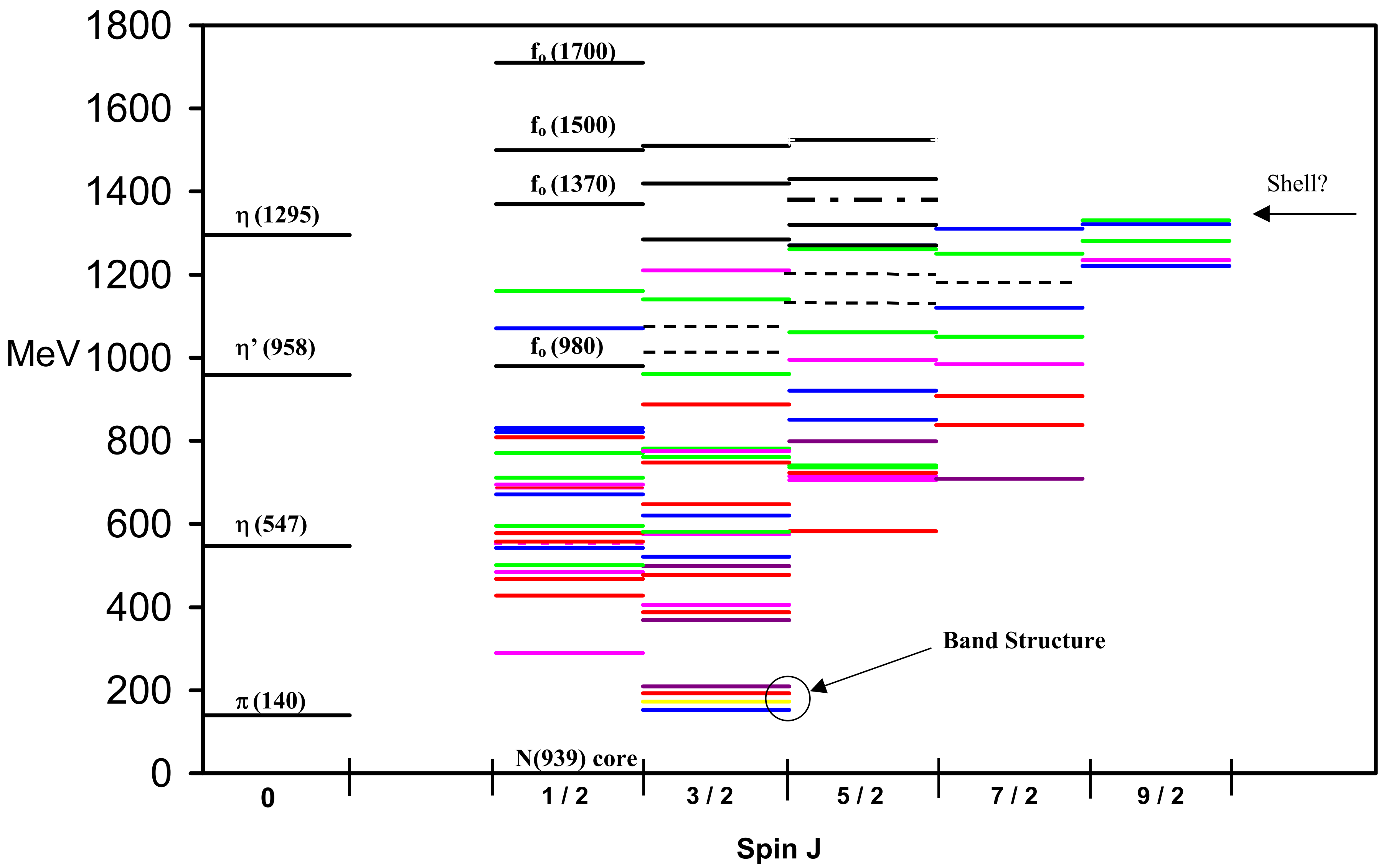

Fig. 5.